\documentclass[10pt, conference, letterpaper]{IEEEtran}
\IEEEoverridecommandlockouts

\usepackage{import}
\usepackage{packages_infocom}
\usepackage{acroRss_infocom}
\setlist[itemize]{leftmargin=*}

\def\BibTeX{{\rm B\kern-.05em{\sc i\kern-.025em b}\kern-.08em
    T\kern-.1667em\lower.7ex\hbox{E}\kern-.125emX}}

\begin{document}

\title{WiLoc: Massive Measured Dataset of Wi-Fi Channel State Information with Application to  Machine-Learning Based Localization\\
}

\author{
Yuning Zhang$^*$, 
Lei Chu$^*$, 
Omer Gokalp Serbetci$^*$, 
Jorge Gomez-Ponce$^{*}$\textsuperscript{\textdagger},
and Andreas F. Molisch$^*$ \\

{\small $^*$ University of Southern California, Los Angeles, CA, USA}; 
{\small \textsuperscript{\textdagger} ESPOL Polytechnic University, Guayaquil, Ecuador, ~~}

}


\maketitle

\begin{abstract}
    Localization is a key component of the wireless ecosystem. \Ac{ML}-based localization using \ac{CSI} is one of the most popular methods for achieving high-accuracy localization with low cost. However, to be accurate and robust, ML-based algorithms need to be trained and tested with large amounts of data, covering not only many \ac{UE}/target locations, but also many different \acp{AP} locations to which the \acp{UE} connect, in a variety of different environment types. This paper presents a massive-sized \ac{CSI} dataset, WiLoc (Wi-Fi Localization), and makes it publicly available. WiLoc is 
    obtained by a series of precision measurement campaigns that span three months, and it is massive in all the above-mentioned three dimensions:  $>12$ million UE locations, $>3,000$ \acp{AP}, covering $16$ buildings for indoor localization, and $>30$ streets for outdoor use. The paper describes the dataset structure, measurement environments, measurement protocols, and the dataset validations. Comprehensive case studies validate the advantages of large datasets in ML-driven localization strategies for both ``standard'' and transfer learning. We envision this dataset, which is by far the largest of its kind, to become a standard resource for researchers in the field of ML-based localization.  
\end{abstract}

\section{Introduction}       \label{sec:intro}
    \subsection{Motivation}
        This paper provides and describes a new resource for research in \ac{ML}-based localization using (frequency-resolved) \ac{CSI} between wireless \acp{TX} and \acp{RX}. Specifically, we performed an extensive precision measurement campaign to acquire \emph{labeled} \ac{CSI} data in a number of indoor and outdoor locations, and show that indeed such large amounts of data allow the training and testing of better ML localization algorithms. The database, which is orders of magnitude larger than any comparable one, is being made publicly available with this paper. 
                        
        \textbf{Why ML-based localization?} Wireless localization has become, over the past 30 years, a key requirement for wireless systems due to its numerous applications ranging from asset tracking to various \ac{IoT} applications to robust beamforming\cite{huang2018location,zekavat2019handbook,molisch2023wireless}. Localization can be achieved by multiple methods, such as proximity-based and time-of-arrival/direction-of-arrival-based methods. However, the former ones cannot provide precise coordinates of the device but rather identifies a \emph{region} around a reference point, and the latter ones, which include GPS and cellular E-911 localization, have compromised performance in street canyons, under canopies, and in indoor environments. Furthermore, they require expensive hardware for precision clock synchronization and calibrated arrays for accurate timing/angular determination, respectively \cite{zekavat2019handbook}. For these reasons, \ac{ML}-based localization has recently received growing interest \cite{burghal2020comprehensive,singh2021machine,roy2021survey,yapar2023real}.

        \textbf{Data types for ML-based localization.} ML localization algorithms can be classified according to the data type (physical information) that they use as input \cite{burghal2020comprehensive}. \ac{RSSI} is the most common data type due to the ease of collection. 
        However, \ac{RSSI} can be an ambiguous measure due to fading: the same \ac{RSSI} can occur at multiple locations. Such ambiguity makes the \ac{RSSI}-based learning models sensitive to observation errors and sparsity of observations, \textcolor{black}{although, in some cases, \ac{RSSI} is preferred for its simplicity and reduced load on processors due to less data. }. Instead, using \acp{CSI} is often preferable since \acp{CSI} provide
        fine-grained amplitude data across the subcarriers, resulting in more accurate localization algorithms\footnote{Observations of the \ac{CSI} at multiple antenna elements of a \ac{UE} or \ac{AP} (also used by \ac{DoA} estimation) can further enhance positioning accuracy, but are beyond the scope of the current paper.}. 
        Additionally, thanks to the use of \ac{OFDM} in Wi-Fi, most of \acp{UE} are capable of capturing \acp{CSI} as part of their normal operation, as it is needed for demodulation \cite[Chap. 15]{molisch2023wireless}. 
        
        \textbf{The need for large measured CSI datasets.} By its very nature, \ac{ML} is predicated on the availability of large datasets. This applies, separately or together, to the training and the testing. Even ML algorithms designed to operate with a small amount of training data need to undergo testing in a variety of environments, for which large amounts of realistic data need to be collected.   
        
        Thus, massive realistic datasets are always needed for developing and testing better and/or more robust localization algorithms, where the word ``massive'' specifies the dimensions \emph{(i) number of \acp{UE} locations, (ii) number of \acp{AP}, and (iii) number of different environments} that have been covered, such as different buildings and streets. This creates the diversity of data characteristics that is required to avoid learning only location-specific features and thus run into overfitting problems. Moreover, with more locations, datasets can provide improved scalability for transfer learning across different environments. Current publicly available measured datasets show severe limitations in at least one of the above-mentioned dimensions, as will be discussed in more detail in Sec. \ref{sec:SoTA}.   
        Synthetic data can be generated by ray-tracing, but the accuracy may suffer from the imperfect object location and material modeling in databases, as well as insufficient modeling of diffraction and diffuse scattering\footnote{While these problems can be partly overcome by ray tracing with lidar-scan based point cloud representations of the environment, the effort in collecting and processing scans over a large area and/or large number of buildings and running ray tracing over them is \textcolor{black}{costly; a third-party \ac{LiDAR} scanning service for an office building can be on the order of \$$10$k at the time of this writing in the Los Angeles area.}} which in turn may result in suboptimum algorithm design and/or misleading performance assessment. 
        
        We note that ML-based localization algorithms are facing a variety of challenges, which are discussed at length in a numerous survey and research papers. The goal of the current paper is not to overcome all those challenges. Rather, we make the point that ML-based localization is already established as a promising and very active research area with hundreds of publications, and any realistic assessment, and further progress in it, requires extensive and realistic data sets.      
        As a result, \emph{a massive real-world {\em measured} \ac{CSI} dataset with numerous \ac{UE} locations, \acp{AP}, and environments is urgently needed by the \ac{ML} community.}

    \subsection{Challenges and solutions:} 
        To fulfill this need, this paper presents WiLoc, the by-far largest \emph{publicly available} database of labeled \ac{CSI} data, i.e., \acp{CSI} and associated locations and \ac{AP} information. WiLoc has more than $12$ million \ac{UE} locations and $3,000$ \acp{AP},  covering 16 buildings with different floor plans and floor counts, and over $30$ outdoor streets with various widths and lengths, indicating its massive size in \emph{ALL} three dimensions mentioned above. 
        
        We resolved three fundamental problems to enable the data collection for WiLoc: (i) how to deploy thousands of \acp{AP} over a large area; (ii) how to precisely measure \acp{CSI} on different frequencies simultaneously, i.e., not miss any \ac{CSI} due to channel sweeping, and (iii) how to accurately obtain labels, i.e., the location and \ac{AP} information associated with a particular \ac{CSI}. 
        
        \textbf{(i) Large-scale deployment of \acp{TX}.} Deployment of large numbers of dedicated measurement \acp{TX} is practically impossible. Thus the first challenge is solved by using the existing Wi-Fi network in \textcolor{black}{the \ac{USC}} campus with thousands of Wi-Fi \acp{AP}\footnote{We have consulted with the university IT department and followed all their regulations during the measurements and the publication of the dataset. }. This method simultaneously ensures (a) having a massive number of \acp{AP} in different environments, (b) having a realistic deployment, and (c) avoiding interference to/from an existing network. Since we do not have the control of the \acp{AP}, to avoid the problem of unpredictable transmission times and potential beamforming distortions, we acquire the \ac{CSI} only from beacon frames, which are transmitted at regular intervals without beamforming, as enforced by the IEEE 802.11 standard\cite{2025ieee80211}. 
        
        
        \textbf{(ii) Precision \ac{RX}.} The second challenge is the monitoring of transmissions on multiple transmission frequencies. The campus \acp{AP} are configured to transmit on one of three Wi-Fi channels (called 1, 6, and 11) in $2.4$ GHz \ac{ISM} band. Typical Wi-Fi data cards or sniffer programs scan the channels sequentially, thus potentially missing transmissions. Furthermore, such devices are typically consumer-grade electronics with large distortions, unpredictable antenna patterns, and significant temperature drift. We used a \ac{USRP}, a type of \acp{SDR}, at the \ac{UE} side that digitizes and records the signals over the whole $100$ MHz bandwidth of the \ac{ISM} band. This allows continuous all-channel \ac{CSI} capturing for all \acp{AP}. Extensive verification experiments validated the superior accuracy and stability of this setup compared to sniffers or Wi-Fi cards.  
        
        \textbf{(iii) Accurate labeling.} Accurate labels are obtained by a combination of a self-built measurement cart with an optical encoder synchronized to the \textcolor{black}{\ac{RF}} equipment, and measuring on carefully laid-out tracks. 
        
        \textbf{Verification} All the above measures ensure high accuracy for the acquired \ac{CSI}; residual errors due to drift of equipment components and human movement in the environment are quantified as well. Note that it is possible to emulate amplitude quantization errors, frequency drift, and other impairments of consumer-grade equipment by adding such distortions in postprocessing to our precision measurements, while the converse (measuring with low-precision equipment and concluding the true CSI properties) would not be possible. 
    \subsection{Contribution of the paper}
        To summarize, the main contributions of this paper are:
        \begin{itemize}[]
            \item We present WiLoc, a massive realistic \ac{CSI} dataset for ML-based localization with $>3,000$ \acp{AP}, $>12$ million of \ac{UE} locations, $180$ indoor trajectories from $16$ buildings, and $>30$ outdoor streets in an urban scenarios, which is orders of magnitude larger than any comparable existing dataset. {\em As part of this paper, we make this dataset publicly available.} 
            \item We describe in detail the measurement procedures, as well as evaluation and validation steps.
            \item We demonstrate examples of the impact of the dataset size and diversity of training data on existing localization algorithms, covering both traditional training and transfer learning. 
        \end{itemize}

        We emphasize again that the purpose of this paper is to present the details of the WiLoc dataset and an analysis of the importance of large datasets for algorithm training and testing. We do {\em not} aim to present here new ML localization algorithms, but rather provide a basis on which researchers in this community can develop and test such algorithms. 
    \begin{table*}[!t]   
        \centering
        \vspace{12 pt}
        \caption{CSI-available Datasets. All datasets used a single RX to measure at all the UE locations.}
            \begin{tabular}{|c|c|c|c|c|c|c|c|c|}
                \hline
                \textbf{Ref.} & \textbf{Year} & \textbf{Dataset} & \textbf{AP} & \textbf{UE} & \textbf{Coverage} & \textbf{Environment} & \textbf{SR/AP}\\ \hline \hline
                \cite{ayyalasomayajula2020deep} & 2020 & WILD & $4 \sim 8$ & 108,008 & 47 m$^2$/139 m$^2$ & 1 B, 2 corridors & 20 Hz \\ \hline
                \cite{arun2022p2slam} & 2022 & P2SLAM & $5 \sim 10$ & 105,639 & 750 m$^2$/2000 m$^2$ & \makecell{1 B, 10 corridors \& \\ 3 rooms} & 20 Hz \\ \hline
                \cite{cominelli2022antisense} & 2022 & CSI-based... & 2 & 8 & 46.2 m$^2$ & 1 B, 1 room & N/A \\ \hline
                \cite{wang2022framework} & 2022 & CSU & 6 & 347 & \makecell{3608 m$^2$ \\ (length: 183.3 m)} & 1 B, 7 corridors & N/A \\ \hline
                \cite{hanna2022wisig} & 2022 & WiSig & 174 & 41 & 400 m$^2$ & 1B, 1 room & N/A \\ \hline
                \cite{burmeister2022highresolution} & 2022 & High-resolution... & 1 & 87,500 & length: 70 m & 1 B, 1 corridor & 1 kHz \\ \hline
                \cite{strohmayer2024wifi} & 2024 & HALOC & 1 & 138,885 & \makecell{52 m$^2$ (20 m)} & 1 B, 1 corridor & 100 Hz \\ \hline
                \textbf{\makecell{WiLoc \\(Indoor)}} & \textbf{2025} & \textbf{WiLoc} & \textbf{3293} & \textcolor{black}{$\mathbf{6,029,581}$} & \textbf{length: 4560 m} & \textbf{16 B, 180 corridors} & 10 Hz \\ \hline
                \textbf{\makecell{WiLoc \\(Outdoor)}} & \textbf{2025} & \textbf{WiLoc} & \textbf{24} & \textcolor{black}{$\mathbf{6,555,588}$} & \makecell{\textbf{450,000 m}$\mathbf{^2}$ \\ \textbf{(length: 5263 m)}}  & $\mathbf{>30}$ \textbf{streets} & 10 Hz \\ \hline
            \end{tabular}
        \label{table:datasetComp}
    \end{table*}

    \vspace{-.5em}
    \subsection{Paper organization}
        The rest of the paper is organized as follows: after a description of the state of the art in Sec. \ref{sec:SoTA}, the dataset structure and example are described in Sec. \ref{sec:dataset}. This is followed in Sec. \ref{sec:system} by a description of the channel sounding system and environment. The dataset evaluation procedures and validations are given in Sec. \ref{sec:evaluation}. Sec. \ref{sec:mlresults} shows the impact of the dataset size on supervised learning with both training/testing in the same environment, and with transfer learning. After a discussion of the limitations of our data in Sec. \ref{sec:limit}, a summary concludes the paper in Sec. \ref{sec:conclusion}. 
\section{State of the art}      \label{sec:SoTA}
    \textbf{ML algorithms using \ac{CSI}.} While the goal of this paper is not the introduction of new algorithms, CSI-based ML localization algorithms determine the requirements for, and application of, the data. Due to space constraints, and in light of the more than 100 papers published on this topic, a systematic survey of this field is not possible here. Rather, we refer to the several survey papers on indoor and/or Wi-Fi localization that include discussions of CSI-based methods, and the associated references, e.g., \cite{ma2019wifi,zafari2019survey,zhu2020indoor, burghal2020comprehensive,yang2021survey, singh2021machine,roy2021survey,yapar2023real,yadav2023systematic,lin2024state,kerdjidj2024uncovering,sonny2024survey,song2019novel,njima2022dnnbased,abbas2019wideep,sun2014wifi,jang2018indoor}.

    \textbf{RSSI data sets.} We next survey measured data sets for ML or fingerprinting-based localization, in particular those that are publicly available. The large majority of those provide \ac{RSSI} only. Since their focus is on a less detailed channel representation than the \ac{CSI} at the center of this paper, we only refer to their comprehensive survey in \cite{feng2024review}. We do note, however, that almost all the surveyed campaigns used non-precision \acp{RX}, such as Wi-Fi cards or smartphones (see Sec. 3 for a discussion of that approach), and are limited to one or two buildings (with the exception of \cite{ashraf2022wi}, which has six buildings, but whose emphasis was on the impact of different users and different types of smartphones).

    \textbf{CSI data sets.} On the other hand, there are only a few publicly available datasets that provide \acp{CSI}, as listed in Table \ref{table:datasetComp}. It is worth noting that all of them (except the proposed one) are indoor-only. 
    The ``AP'' and ``UE'' columns represent the amount of available \acp{AP} and \ac{UE} locations in each dataset. The coverage represents the area of the convex hull (or circumscribed rectangle) of the region\footnote{Usually, the square edges are parallel to, either the trajectories of a moving \ac{UE}, or the track of a UE that is deployed to multiple locations. } that includes all the UE locations. To save space, we use ``$X$ B'' to denote the number of buildings the dataset covers; all datasets, with the exception of ours, are from one building only. We classify hallways as corridors, as they are similar indoor structures by nature. Rooms can be living room-style spaces and/or offices and/or laboratories. The SR/AP column represents the location sampling rate \emph{per AP}. which describes the repetition rate in Hz at which the \acp{CSI} from the same \ac{AP} are collected; this rate can be mapped to the distance domain combined with the \ac{UE} movement speed, which is typically on the order of $0.5-1$ m/s. Please note, with more \acp{AP}, the distance between adjacent fingerprints (although they may contain \acp{CSI} from different \acp{AP}) may also get denser. The proposed dataset has a high rate of detecting \acp{CSI} from different \acp{AP} due to continuous \ac{CSI} capturing. Ref. \cite{arun2022p2slam,ayyalasomayajula2020deep} installed a Wi-Fi transceiver on a robot platform that can precisely track locations. In contrast, Ref. \cite{strohmayer2024wifi} uses human-carried equipment along with a walking capture along a zig-zag path in a corridor. Ref. \cite{cominelli2022antisense,wang2022framework,hanna2022wisig} used fixed \ac{UE} locations at which to acquire the \acp{CSI}, so that the repetition rate in Hz is not applicable; the distances between the measurement points are $\sim2$m, $<1$m, and $1$m, respectively. Particularly, \cite{cominelli2022antisense} performed $2$ temporal stability verification but only with 1 second each; \cite{hanna2022wisig} used the Orbit grid testbed, which contains a static 20x20 node grid (each node has a stand-alone \ac{USRP} that can be configured as either \ac{TX} or \ac{RX}) in an empty indoor environment; \cite{hanna2022wisig} took $174$ nodes as \acp{AP} and selected other 41 \acp{USRP} as \ac{RX}. However, all nodes were static all the time, and the indoor environment lacked typical indoor structures between \acp{AP} and \acp{UE}, such as walls, corners, and doors. Ref. \cite{burmeister2022highresolution} presents a precise dataset, however, with limited locations. Additionally, the \ac{AP} is a \ac{USRP} instead of a commercial Wi-Fi \ac{AP}. 

    To summarize, all the datasets above are valuable for the \ac{ML} community; however, they are limited in one or more dimensions we mentioned earlier. Particularly, our proposed dataset is larger by  $10$x and $30$x in the number of \acp{UE} and \acp{AP}, respectively. Additionally, we covered $16$ different buildings with $180$ corridors, which is unique among all data sets. Finally, our data set is the only one with outdoor \acp{UE} locations; these provide \acp{CSI} to both indoor and outdoor \acp{AP} that cover about $450,000$ m$^2$ with more than $30$ streets. 
    Thus, WiLoc is the by far most extensive available dataset.  
\section{Dataset Overview}  \label{sec:dataset}

    \subsection{Dataset Statistics}
    As previously mentioned, the proposed dataset has a huge number of available \ac{UE} and \acp{AP} locations, as well as various indoor and outdoor environments. Obviously, not all \ac{UE} locations can ''see" all $3000$ \acp{AP} simultaneously. Table \ref{table:dataset_stat}\footnote{\textcolor{black}{Due to through-wall/window propagation, direct summation of indoor/outdoor \acp{AP} in indoor/outdoor data may exceed the total corresponding \ac{AP} amount. }} provides a high-level summary of the statistics of the detectable number of \acp{AP} in the different environments. 
    Users can fully customize their methods of interpolation or binning of the sample locations. 
     \begin{table} 
        \centering
        \vspace{12 pt}
        \caption{Key dataset characteristics}
        \begin{tabular}{|c|c|}
            \hline
            \textbf{Parameter} & \textbf{Value} \\ \hline \hline
            Average Indoor AP Amount per building & $126.19$  \\ \hline
            Average Outdoor AP Amount per building & $1.68$  \\ \hline
            Indoor APs in Indoor Data & 1417  \\ \hline
            Indoor APs in Outdoor Data & 2739  \\ \hline
            Outdoor APs in Indoor Data & 10  \\ \hline
            Outdoor APs in Outdoor Data & 24  \\ \hline
        \end{tabular}
        \label{table:dataset_stat}
    \end{table}

\begin{figure}
        \centering
        \includegraphics[width=1\linewidth]{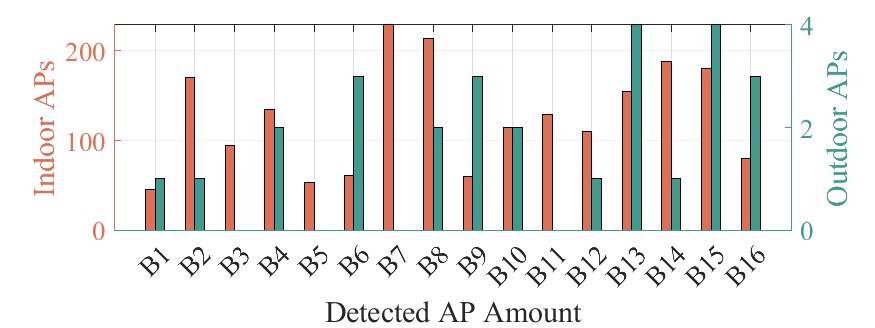}
        \caption{\small{The detected number of APs across all 16 buildings. The left Y-axis (left half pair) indicates the visible indoor APs, and the right Y-axis (right half pair) indicates the visible outdoor APs. }}
        \label{fig:visibleAPAmountPerBuilding}
    \end{figure}

    Moreover, Fig. \ref{fig:visibleAPAmountPerBuilding} shows the distribution of the visible \acp{AP} for each numbered building (note that an indoor \ac{AP} visible in a building does not need to be located in that building). The building indices are denoted as $B \in \{0,1,2,...,16\}$, where $B=0$ represents the outdoor scenario. 

    \subsection{Dataset Metadata}  \label{subsec:dataset_metadata}
         Each building, as well as the outdoor dataset, contains a metadata file and a \ac{CSI} file. All \acp{CSI} have an associated set of metadata with the following $14$ assets: channel index
         , relative timestamp, relative distance, cart speed, cart heading indicator\footnote{ We define $1$ as the ``+X'' direction, then increase with every $0.5$ for every $45^\circ$ counterclockwise bearing. }, \ac{AP} label\footnote{The \ac{AP} label is defined in the format of ``$A+\frac{N}{10}$'', where $A \in \mathbb{Z}$ represents physical \acp{AP}, and $N\in\{0,1,2\}$ stands for $3$ campus-wise networks that are configured in the same \ac{AP}. An \ac{AP} label can be uniquely mapped to a \ac{BSSID}, or a \ac{MAC} address, which we do not make accessible to the public in the dataset due to IT security policy. }, $X$, $Y$, $F$, \ac{RSSI}, noise and interference, \ac{SINR}, repetition index\footnote{For building $B2$, each line was measured $4$ times, over several days. }, and measurement index\footnote{As described in Sec. \ref{subsec:mobility}, each measurement campaign follows a straight line, and is denoted with a distinct index. Repeated measurements on the same line also have different indices. In this paper, the words "trajectory" and "line" are interchangeable. }. The building index $B$ can be retrieved directly from the dataset name itself. We define $F=1$ for the outdoor dataset. Table \ref{table:metadata} shows the detailed description of each asset. 
         \begin{table}     
            \centering
            \vspace{12 pt}
            \caption{Metadata elements (per \ac{CSI})}
            \resizebox{\linewidth}{!}{
                \begin{tabular}{|c|c|c|}
                    \hline
                    \textbf{Metadata} & \textbf{Description}\\ \hline \hline
                    Channel Index & Any value takes from the set $\{1, 6, 11\}$ \\ \hline
                    Timestamp & \makecell{Per-line based; \\ the time elapse from the starting point; in seconds}   \\ \hline
                    Distance & \makecell{Per-line based; \\ the distance from the starting point; in meters} \\ \hline
                    Speed & Instant cart movement speed, in m/s  \\ \hline
                    Heading & Indicates the direction that the cart moves  \\ \hline
                    \ac{AP} Label & Indicates the physical \acp{AP} and subnetworks  \\ \hline
                    $X$ & The $X$ local coordinate of the \ac{CSI}, in meters \\ \hline
                    $Y$ & The $Y$ local coordinate of the \ac{CSI}, in meters \\ \hline
                    $F$ & The floor index in a building \\ \hline
                    \ac{RSSI} & \ac{RSSI} at location $(X,Y,F,B)$, in dBm \\ \hline
                    N \& I & Noise \& interference at $(X,Y,F,B)$, in dBm \\ \hline
                    \ac{SINR} & \ac{SINR} at location $(X,Y,F,B)$, in dB \\ \hline
                    Repetition & The repetition index of a line being re-measured \\ \hline
                    Measurement & The global index of each measured line \\ \hline
                \end{tabular}
            }
            \label{table:metadata}
        \end{table}

    \subsection{Example results}
        Here we show examples of \acp{RSSI} (because of the ease of visualizing them as a function of location) and full \ac{CSI}. The data were captured from $B=2, F=5$, network $N=0$, and measurement index $\in \{1, 5, 9, 13, 17, 21\}$. 
        
\begin{figure}
            \centering
            \includegraphics[width=.8\linewidth]{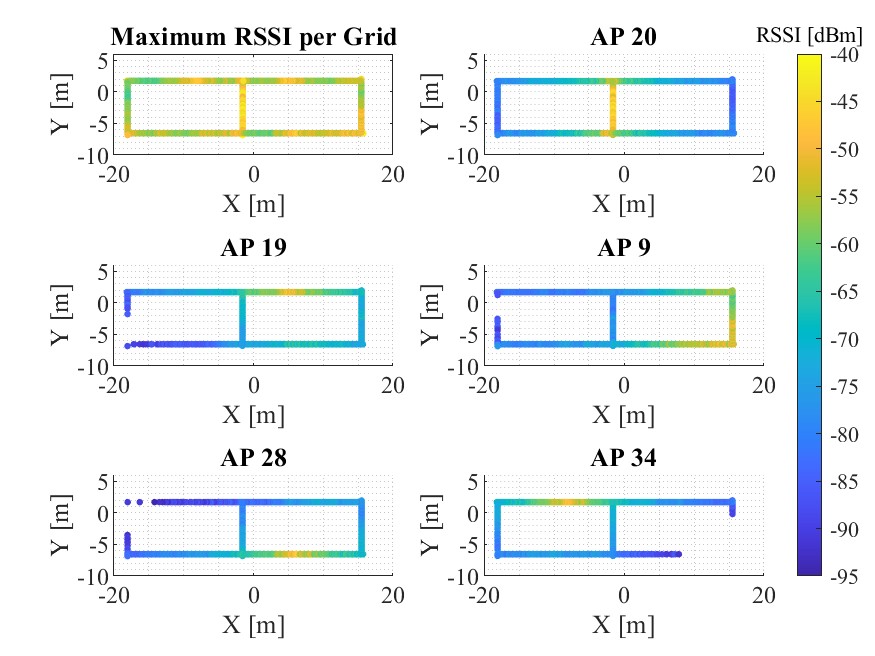}
            \caption{\small{RSSIs from different APs along the same UE trajectory (on floor 5 of Building 2). }}
            \label{fig:samefloor}
        \end{figure}

        Fig. \ref{fig:samefloor} shows plots of the RSSI from different \acp{AP} along the selected measurements. Each pixel\footnote{\textcolor{black}{Pixels are too dense to be separable visually. }} in the plot corresponds to the average \ac{RSSI} in a $0.1$-m sidelength square. The sub-figures titled by ``AP $A$'' show the \acp{RSSI} from five strong \acp{AP} along the lines; note, however, that more than these five \acp{AP} might be visible at each location of the \ac{UE}. The top left figure shows the results when, in each square along the trajectory, the strongest \ac{AP} is selected.  

\begin{figure}
            \centering
            \includegraphics[width=.8\linewidth]{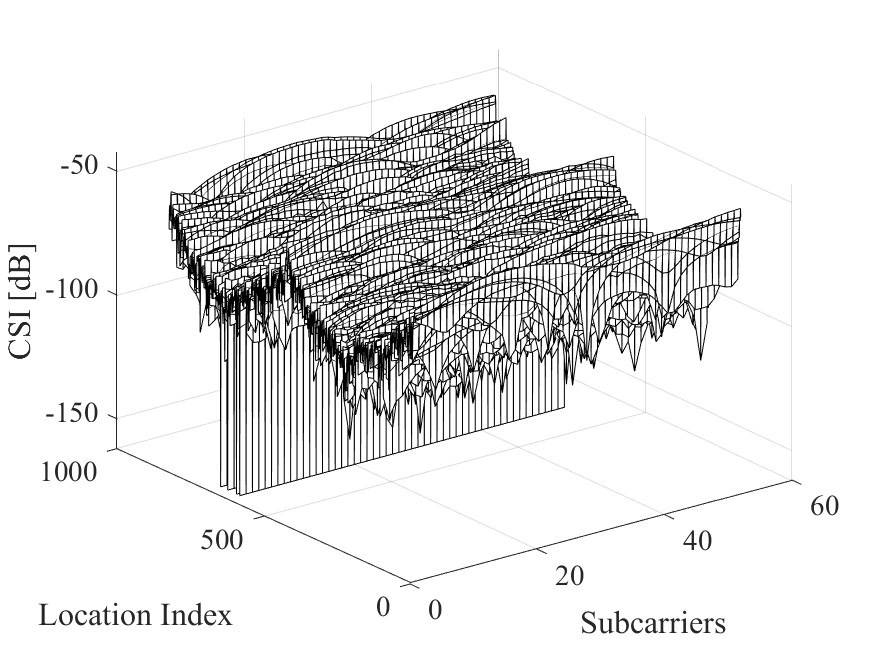}
            \caption{\small{CSI evolution over the trajectory without the central corridor - AP 19. }}
            \label{fig:csiTraj}
        \end{figure}

        Fig. \ref{fig:csiTraj} shows the evolution of the CSI magnitude along the outer trajectory (without the middle line), of \ac{AP} $19$. The location index $=0$ represents the maximum \ac{RSSI} location in Fig. \ref{fig:samefloor} at $(X,Y)=(6,3)$, and the \ac{UE} moves counter-clockwise and ends at the start location. Locations where  \ac{AP} $19$ cannot be detected are replaced by $-160$ dB across all subcarriers, for presentation purposes. 

        We discuss more details of the measurement fundamentals and data collection principles in the next section.

\section{Measurement System and Procedure}     \label{sec:system}
    This section presents the measurement setup and procedures. This information is important for assessing the details and the quality of the acquired data. Various verification measurements and sanity checks are described in Sec. \ref{sec:evaluation}. 
    
    Our measurement system consists of a large set of operative Wi-Fi \acp{AP}, and a \ac{USRP} 
    to collect \acp{CSI} to different \acp{AP}. With a \ac{USRP} inaccuracies introduced by thermal drift of the component characteristics are minimized compared with various off-the-shelf Wi-Fi receivers. Additionally, \acp{USRP} usually have wider \ac{BW} for multi-channel measurements, instead of sequential channel switching common for channel sniffers. In general, \acp{SDR} can provide more precise and accurate results, while retaining the ability to emulate consumer products by adding errors in postprocessing as discussed in Sec. I.B. Similarly, missing beacons because of sequential scanning, which affects commercial \acp{UE} and sniffers, can be easily implemented in postprocessing based on the complete data availalbe in WiLoc.
    
    Since the \acp{CSI}-associated \ac{UE} locations must be known exactly, the \ac{RX}\footnote{We use the words ``\ac{RX}'' and ``\ac{UE}'' interchangeably. } must be installed on a cart with the capability to accurately track and record locations.

\vspace{-.5em}
    \subsection{Sounder Structure}   \label{subsec:sounder}
        \subsubsection{Transmitter}   \label{subsubsec:tx}
\begin{figure}
                \centering
                \includegraphics[width=.7\linewidth]{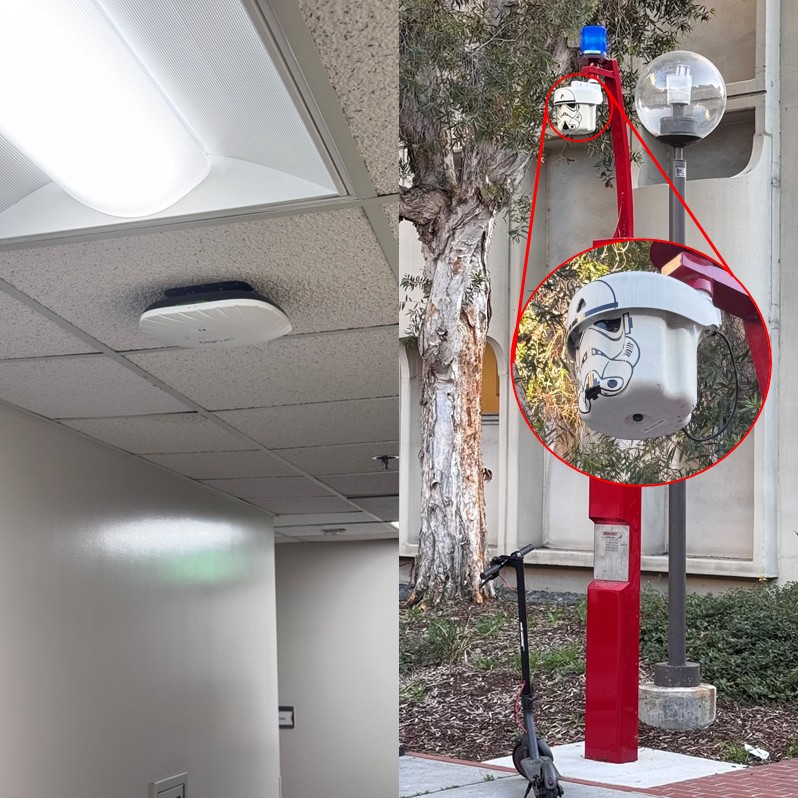}
                \caption{\small{Example photos of indoor (left) and outdoor (right) APs deployed across the campus. }}
                \label{fig:AP}
            \end{figure}

            There are two different types of Wi-Fi \acp{AP} across the entire campus, for indoor (on ceilings) and outdoor (on poles and side walls) use, respectively; they differ in their deployment as well as housing and \ac{RF} characteristics. Fig. \ref{fig:AP} shows photos of each type. Since most indoor \acp{AP} are installed inside access-controlled rooms, indoor footprints were captured in corridors only; consequently, most of the in-room \acp{AP} do not have \ac{LOS} to the \ac{RX}. Most of the outdoor \acp{AP} are $3\sim 4$ m above the ground. 
            
            The three non-overlapping channels are used to prevent inter-channel interference. Neighboring \acp{AP} usually operate on different channels. However, all three networks on the same \ac{AP} use the same channel. Since the propagation channels are unique to the \ac{AP} but do not depend on the network, \acp{CSI} obtained from different beacons of the networks on one \ac{AP} are highly correlated, especially since these beacons are transmitted within a short time (order of $1$ms) from each other. The redundancy of these networks can be used for noise averaging or assessment of short-term variations. 
            
            We stress that the proposed dataset provides \emph{raw measured \acp{CSI}}, without compensation for \ac{RF} distortions in the \acp{AP} due to policy restriction and practical requirements\footnote{Usually a calibration measurement for one \ac{AP} takes more than one hour, and our dataset involves thousands of \acp{AP}. Furthermore, disconnecting \acp{AP} to measure them in an anechoic chamber was not allowed by the IT department. }. Transmissions from calibrated precision \acp{TX} such as used in \cite{burmeister2022quantifying} provides higher accuracy, but it is practically impossible and cost-prohibitive to deploy a massive number of such \acp{TX}.

        \subsubsection{Receiver}   \label{subsubsec:rx}
            At the \ac{RX}, a nominally omni-directional antenna is installed on the top of the \ac{RX} cart at $1.7\sim2.1$ m above the ground. 
            A permanent marking on the antenna and the cart ensures that the antenna always has a fixed relative direction with respect to the cart throughout all measurement campaigns. 
            Note when the signals incident from (almost) the top of the antenna, there is an attenuation around $15$ dB. 
            
            The antenna is connected to the \ac{USRP} via two concatenated \acp{BPF}. The \ac{USRP} digitized the signal with $100$ MSamp/sec at \ac{BB}. The movement distance of the cart is captured by an on-wheel optical encoder, see Sec. \ref{subsec:mobility} for more details. The capturing of \ac{RF} signal and encoder data are synchronized to start simultaneously.

\vspace{-.5em}
    \subsection{Mobility}    \label{subsec:mobility}  
\begin{figure}
            \centering
            \includegraphics[width=.7\linewidth]{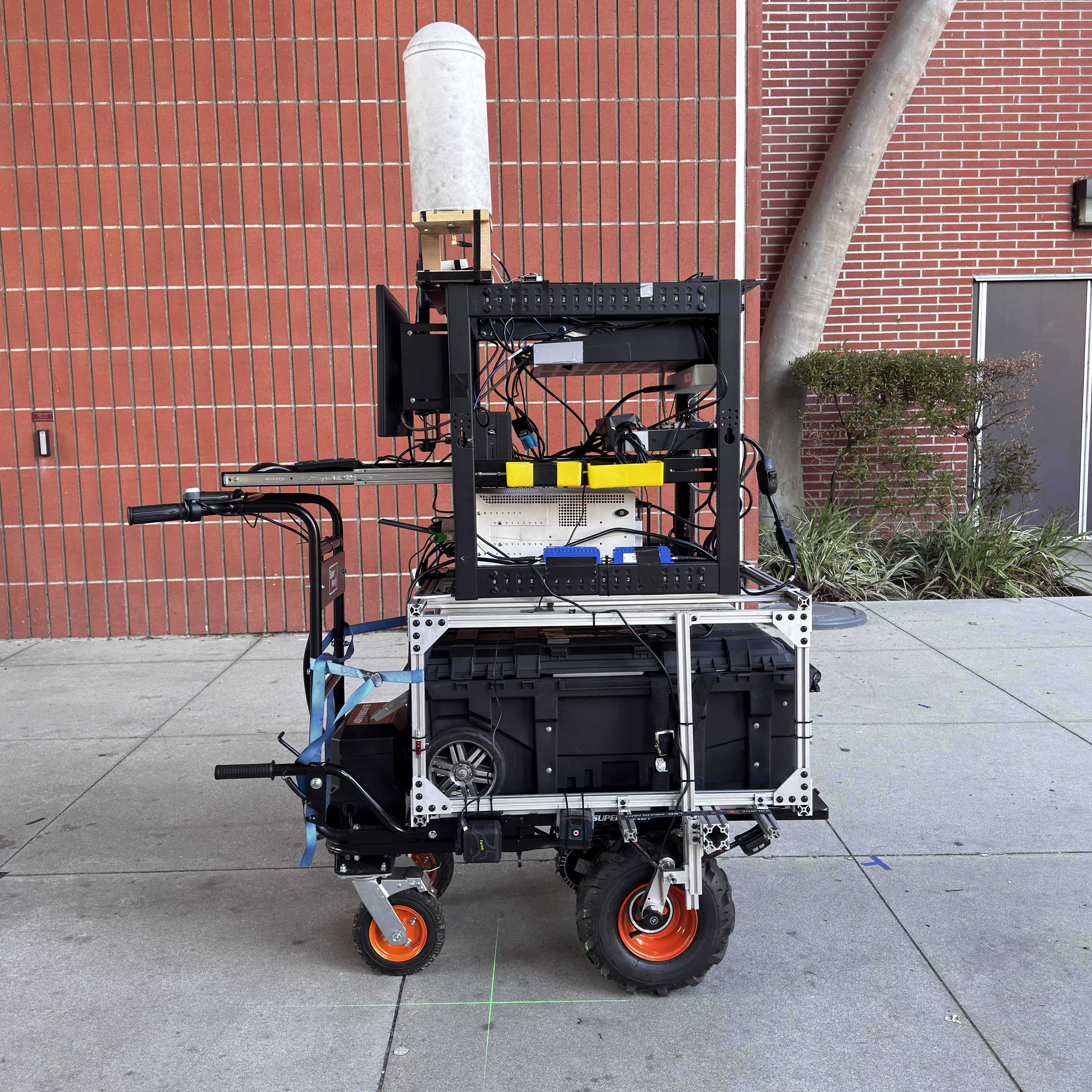}
            \caption{\small{The right-side view of the RX cart, with laser level on. }}
            \label{fig:rx}
        \end{figure}

        Fig. \ref{fig:rx} shows a photo of the \ac{RX} cart, and we moved the cart towards the small-wheel end with an average speed of $0.45$ m/s. In Building 2, the average cart speed was $0.15$ m/s. The details on the cart construction, distance recording, and accuracy assessment will be published along with the dataset due to space limitations here. 
        
        It is vital to record accurate locations as accurate ground truth for the \ac{ML} models. To avoid distance drifts brought by \acp{IMU}, such as the Intel-T265, caused by varying light conditions and different background textures, we used linear tracking with the assistance of a dual-phase optical encoder\footnote{\textcolor{black}{A dual-phase optical encoder can record movement in both directions. It has a resolution of $1.5$ mm and a constant sampling rate of $5$ kSamp/sec. }} to \textcolor{black}{record timestamped cart movement distance}. To achieve $2$-D tracking, we consequently moved the cart only on trajectories that are piecewise linear, with a precisely measured start point coordinate, movement direction, and movement distance. We then map the trajectories to a campus-wide coordinate system for outdoor fingerprints and building-specific coordinate systems for indoor measurements. We used ``track marker'', such as grooves between stone tiles or tape (applied to the ground to form a straight line indicator), and a cross-line laser level, see the green cross on the ground in Fig. \ref{fig:rx}, to ensure the cart moves along straight lines. 
\vspace{-.5em}
    \subsection{Measurement Procedure }    \label{subsec:measurement}
        The measurement campaigns for indoor and outdoor scenarios share the same principle, namely, moving on straight lines. For each measurement, the \ac{RX} started with the alignment to a point marked by a masking tape, as the starting point, and when it approached the ending point, 
        we kept the cart moving with the same speed to go over the ending point for $40\sim60$ cm, then drove it back to get aligned with the ending point. Such a method ensures the fingerprints around the ending point are (approximately) equally spaced
        \footnote{In post-processing, we removed the data for the slowdown and driving-back parts from each line measurement.}.
                   
        The length of the measurement track becomes an issue in outdoor measurements. Outdoor streets can be up to (and in a few cases even exceeding) $500$ m. Since we required the operator to be bowed down when driving the cart to avoid blocking the antenna with their head, while at the same time following the track markers and maintaining a constant speed, measuring a long track in one go would be exhausting and may increase errors. As a result, longer streets were divided into several contiguous shorter lines, and measured individually. For narrow streets/pathways, we only measured once on one side of the sidewalks or in the middle (when it was safe to do so); for wider streets, we measured on both sides. 
        The time the measurement campaigns (including setup, breakdown, etc.) took was approximately 400 hours, distributed over a time period of three months. Between 4 and 7 people were involved in each measurement, including blocking off measurement sites, creating track guides, etc. 
        Measurements both outdoor and indoor were mostly performed during evening/nighttime to minimize the presence of human movement in the measurement environment.

        \begin{figure}
            \centering
            \includegraphics[width=.8\linewidth]{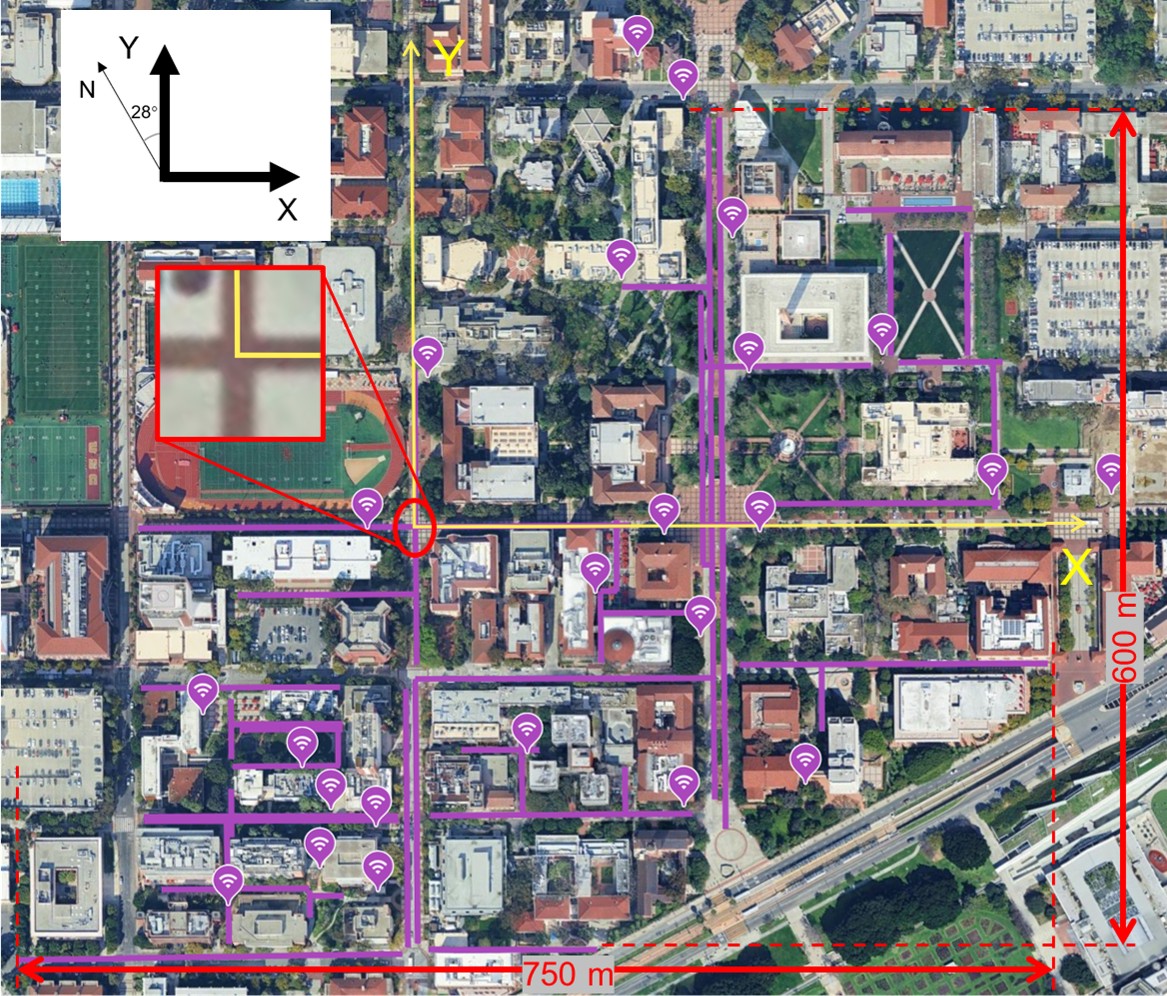}
            \caption{\small{Dataset coverage and measurement map, marked with the boundaries of the outdoor measured locations and the coordinate system axes. }}
            \label{fig:map}
        \end{figure}
\vspace{-.5em}
    \subsection{Measurement environments }    \label{subsec:environment}
        For the indoor environment, only corridors were measured, since we could not get consistent access to the offices. All the $16$ measured buildings have different floor plans, number of floors, number of rooms, and number of \acp{AP}. But most of them share similar structures across different floors. As per university policy, the exact locations of in-room \acp{AP} are confidential. Still, we know that they are all in NLOS to the \ac{RX} on the corridor. The corridor \acp{AP} are labeled to enable selection of in-room or corridor \acp{AP} during evaluations. The corridors are generally $1.5\sim3$ m wide and $10\sim60$ m long, depending on the building structures. We only measured the buildings as well as the floors whose entrance/elevator doors are wide enough to fit the \ac{RX} cart through them. More details about the dimensions of the buildings and related parameters are given in the metadata of our dataset. 
            
        Fig. \ref{fig:map} shows the satellite map of the measurement environment. The street widths vary between $3$ m and $15$ m. We measured on or next to sidewalks. The coordinates of all outdoor \acp{AP} are available, and their locations are illustrated in Fig. \ref{fig:map} by the purple round markers. The purple line segments are outdoor fingerprinting trajectories, and the yellow lines represent the $X$ and $Y$ axes. Indoor coordinate systems share the same $X$ and $Y$ directions as the outdoor ones. 

\section{Dataset Evaluation}  \label{sec:evaluation}
    \subsection{Extraction of CSIs and Location Labels}
        Since channels were captured simultaneously, in post-processing, we then need to (i) separate different channels, (ii) isolate the campus-\ac{AP} beacon signals while discarding all other data packets, and (iii) extract the required information (\ac{CSI}, \ac{BSSID}, and other metadata, see Sec. \ref{sec:dataset}) from each beacon packet. 
        
        For the separation of the different channels, we apply a circular frequency shift to the \ac{BB} data so that the target channel is centered at $0$ Hz. Then, a digital \ac{LPF} is used to suppress all other channels. We next down-sample the channel-selected data to $20$ MSamp/sec to match the single-channel \ac{BW} according to IEEE802.11. Please note that the $20$-MHz channel is not limited by the \ac{RX} system but the \acp{AP}. If wider \ac{BW} channels, namely, $40$ MHz and $80$ MHz, are available at the $2.4$ GHz band, they can also be evaluated.

        We then use the IEEE 802.11-mandated packet structure to find the start of packets by the \ac{L-STF} and \ac{L-LTF}\footnote{It is called ``legacy'' because it is backwards compatible to the IEEE 802.11g standard and thus can serve to identify packets from this and any subsequent 802.11 standards, including 802.11n, 802.11ac (Wi-Fi 5), 802.11ax (Wi-Fi 6), and 802.11be (Wi-Fi 7).}\cite{2025ieee80211}. \acp{L-LTF} can also be used to extract \acp{CSI}. Next, we decoded \ac{MAC} data to acquire \ac{BSSID}, \ac{ESSID}, and the packet type. We first discard all non-beacon packets, then use \acp{ESSID} to filter out and discard all non-campus-\ac{AP} \acp{CSI}. 
        Lastly, we assign the \ac{UE} locations to the extracted \acp{CSI} and \ac{AP} information by timestamp matching. 
        The core components of the WiLoc dataset are readily prepared as ``\ac{CSI} - location - \ac{AP}'' triplets.  

    \subsection{Verification Measurements}
        \subsubsection{Temporal Stability}
            To verify the temporal stability of the measurement system, as well as the impact of moving people on the measured channel characteristics, we put the \ac{RX} at a $1.5$-m wide corridor, with a \ac{LOS} connection to an indoor \ac{AP}, and we used the \acp{CSI} captured at the same location in $4$ scenarios, namely no movement, and 1, 2, or 3 people passing by without blocking the \ac{LOS} \ac{MPC}, over $6$ minutes each.
            
            \begin{figure}
                \centering
                \includegraphics[width=.9\linewidth]{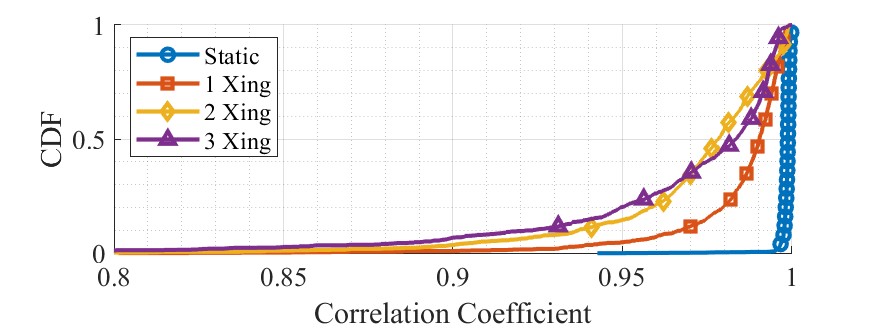}
                \caption{\small{CSI magnitude correlation over time CDF. }}
                \label{fig:indoor_corr_CDF}
            \end{figure}
            
            Fig. \ref{fig:indoor_corr_CDF} shows for all the four scenarios the \acp{CDF} of the cross-correlation coefficients between the magnitude of all \acp{CSI} and the first \ac{CSI}. 
            Fig. \ref{fig:indoor_corr_CDF} indicates that even in a narrow corridor with passing people, the \acp{CSI} can still have at least $95\%$ correlation for $80\%$ of cases. 

            Due to the phase noise of the \acp{AP} and the timing jitter in the transmission of the beacon frame, the phase information of the \ac{CSI} is considerably less reliable, and will change even in a static channel. Thus, while WiLoc provides the complex data, different ML algorithms might decide to use or discard the phase information.

        \subsubsection{Missing beacons}
\begin{figure}
                \centering
                \includegraphics[width=.9\linewidth]{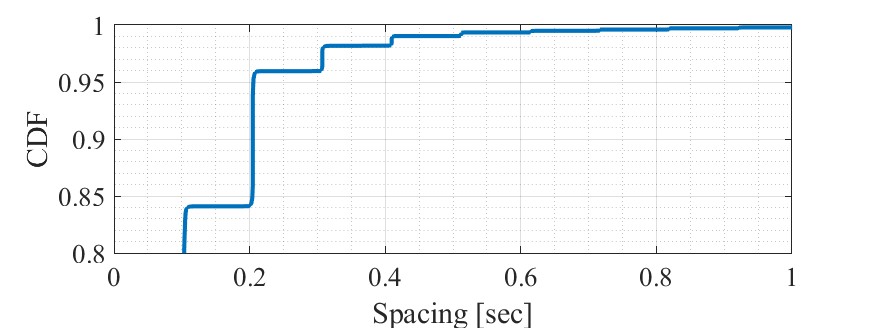}
                \caption{\small{CDF of the beacon spacings in time. }}
                \label{fig:beacon_spacing}
            \end{figure}

            According to the IEEE802.11 specifications, a beacon packet should be transmitted every $0.1024$ sec. 
            However, we observed occasionally missed beacons in the dataset even for fairly good \acp{SNR} ($20\sim30$ dB). To quantify this effect, 
            we selected multiple \ac{RX} trajectories with \ac{LOS} channels, and collected the statistics of the spacings between two consecutive beacons from the same \ac{AP}. Fig. \ref{fig:beacon_spacing} shows the \ac{CDF} of the spacings. It can be observed that in about $85\%$ of cases, no beacon is lost (spacing is the nominal $102.4$ ms), and in less than $5\%$ of cases more than one beacon is lost.
            
\section{Machine Learning Based Localization}   \label{sec:mlresults}
    \subsection{Supervised Learning-Based Indoor Localization}
        In this section, we demonstrate how our collected dataset is utilized in a deep-learning-based indoor localization. As mentioned in Sec. \ref{sec:SoTA}, \emph{the objective of this section is not the development of a new ML-based localization algorithm, but rather provide a baseline, and demonstrate the importance of having a large and diverse dataset for training, testing, and performance comparison}. Our dataset can be used for comparing amongst different ML-based techniques, and even compare them with traditional fingerprinting and proximity-based methods (though the latter approaches obviously would not be suitable for transfer learning).

        

        To showcase the capabilities of the collected dataset and trained deep learning models, we designed a set of experiments and evaluated them under different scenarios. We conducted experiments with various neural network architectures, including transformers and multilayer perceptrons (MLPs). We finally selected a fully connected feed-forward neural network with four hidden layers, each consisting of 512 neurons for the generation of results in this section, due to its simplicity and strong performance. We employed a decreasing learning rate starting from $10^{-4}$ with Adam optimizer. We split the collected dataset into 80\%, 10\%, and 10\% partitions for training, validation, and testing datasets. The models are trained up to 150 epochs with 32 batch sizes. To construct a feature for each bin, we follow the common structure in the literature and stack all the imaginary and real parts of selected CSI(s) into a single real-valued vector, collected across all subcarriers and relevant \acp{AP} \cite{yeli_localization}.\footnote{We determined experimentally that inclusion of the phase information was beneficial, despite the challenges discussed in Sec. V.B. 1. } The output of the model is a 2-D location prediction. We trained the models with MSE loss computed with respect to the ground truth location and prediction. We present our results in terms of the \ac{RMSE} computed over the test set to enhance interpretability.

        \begin{figure}
            \centering
            \includegraphics[width=0.85\linewidth]{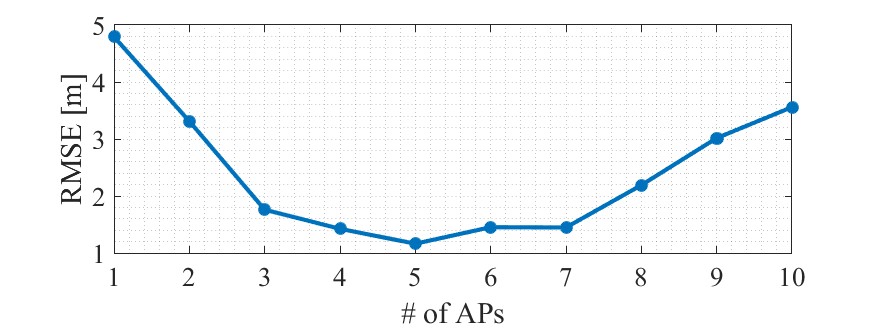}
            \caption{\small{Test set RMSE performance vs No. of APs on $B2,F4$}}
            \label{fig:singleF_FC}
        \end{figure}
        
        For our dataset, we used the floor $4$ of Building 2. The obtained accuracy is comparable to the mean accuracy obtained in \cite{ayyalasomayajula2020deep} and the 90th percentile error in \cite{arun2022p2slam}; it also is considerably better than what could be obtained, e.g., by simple matching to the set of visible \acp{AP} (we tested that to be around $9$ m). In Fig. \ref{fig:singleF_FC}, we present the test set \ac{RMSE} as a function of the number of \acp{AP}. Each data point represents an experiment where all bins utilize the same set of AP(s). It can be observed that increasing the number of APs leads to a significant performance improvement, up to approximately 4 meters, until a certain threshold of five \acp{AP}. The primary reason is that a low number of \acp{AP} (such as one or two) fails to provide sufficient characterization of the bins, resulting in poor performance. As the number of APs increases, the features become more distinctive, enhancing localization accuracy.
    
        However, beyond a certain point, localization accuracy starts to degrade. This outcome is reasonable as there are fewer locations that actually have connections to so many APs, reducing the size of the training set and making it difficult for the model to learn meaningful feature representations. Based on these findings, collecting more data, particularly in challenging \ac{NLOS} environments, is crucial for improving model performance. It is important to note that the results here are particular to this network and that other networks/algorithms might show different behavior. In any case, however, the key is to have more data/APs/buildings. Another point we can conclude from the experiment that the fact that different UE locations ''see" different sets of \acp{AP}, with different sizes, has important impact on performance, changing RMSE by several meters. Such effects, however, are difficult to analyzed with the existing datasets of Sec. II, which in most cases see only $1-4$ \acp{AP}.

\vspace{-.5em}
    \subsection{Transfer Learning in Indoor Localization}
        Although deep learning methods achieve high localization accuracy, they come at a cost of time and labor-intensive data collection. We can mitigate this effort by leveraging a model trained in a data-abundant environment, called the source environment, while the data-scarce environments are considered target environments. This approach is known as transfer learning. We first train a model, $f$, in the source environment, where we have more than 10,000 samples. Then, we fine-tune the pre-trained model in the target environment, which has fewer than 1,000 samples. Fine-tuning is performed with a lower learning rate, as higher learning rates can lead to catastrophic forgetting in such setups. 
        
\begin{figure}
            \centering
            \includegraphics[width=.85\linewidth]{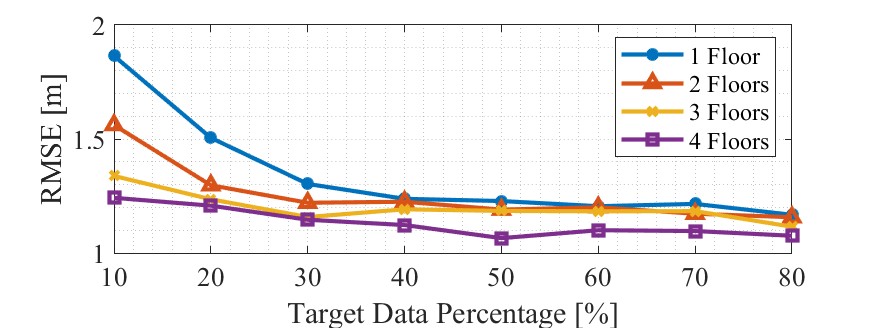}
            \caption{\small{Transfer learning RMSE vs No. of floors in the source dataset and percentage of available data on $B2,F5$}}
            \label{fig:transferlearing}
        \end{figure}

        In Fig. \ref{fig:transferlearing}, we construct the source dataset by combining data from different floors, ranging from the first to the fourth. Then, we fine-tune the model using data collected from the fifth floor. To assess the impact of dataset size in the target environment, we provide results for different percentages of the available data from the fifth floor. The key result presented here is that a model trained on a small source dataset (only one floor) requires more data collection in the target environment, which we aim to minimize. Conversely, by incorporating data from multiple floors and increasing the amount of collected data in the source dataset, the learned representations allow the model to achieve performance with small target datasets that are comparable to cases where smaller source datasets are used with larger target datasets. These effects again demonstrate the importance of large datasets with multiple APs and environments (floors, in this case).
        
\begin{figure}
            \centering
            \includegraphics[width=.85\linewidth]{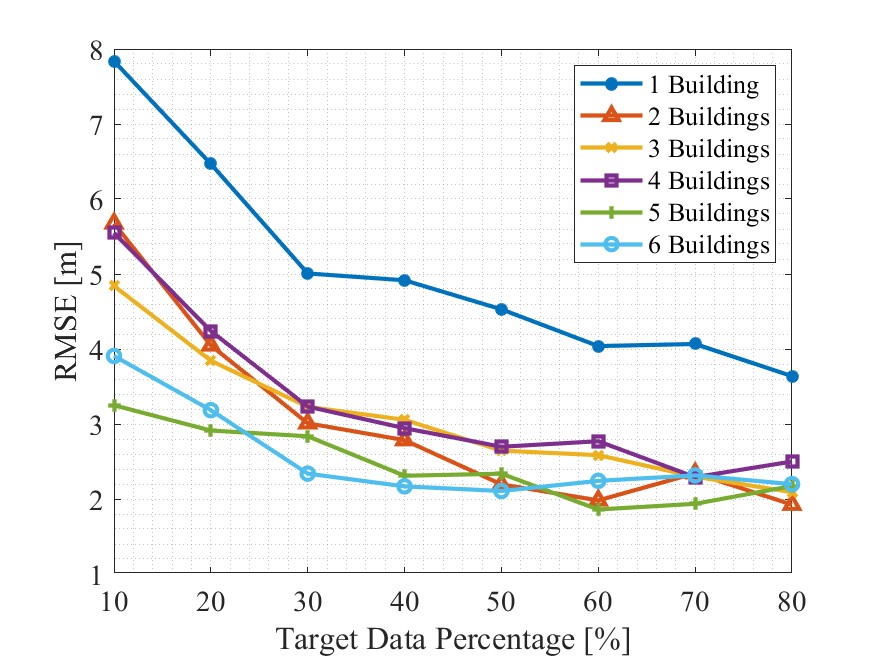}
            \caption{\small{Transfer learning RMSE  vs No. buildings in the source dataset and percentage of available data on $B\textcolor{black}{16},F2$}}
            \label{fig:transferlearingbuilding}
        \end{figure}

        To further emphasize the importance of transfer learning and cross-environment data collection, we conducted an experiment using a cross-building source dataset to pre-train a model, similar to the cross-floor experiment. The source dataset, consisting of second-floor measurements from Buildings 10–1\textcolor{black}{5}, was used to pre-train the model. The pre-trained model was then fine-tuned using data from the second floor of Building 1\textcolor{black}{6}. As shown in Fig. \ref{fig:transferlearingbuilding}, leveraging multiple environments proves to be highly effective compared to using a source dataset with limited diversity. Compared to Fig. \ref{fig:transferlearing}, cross-building transfer learning exhibits significant performance degradation, as learning channel representations are easier within a single building using a cross-floor setup. The primary reason is that a cross-floor setup within the same building shares structural similarities, such as floor plans and materials, which facilitate learning.
        
        However, when the number of source buildings is small, compensating for differences in the \ac{CSI} distribution requires a larger amount of target environment data. This issue becomes even more pronounced in single-building scenarios, where even a large target dataset fails to achieve reasonable localization accuracy. Conversely, when building diversity is high, the model learns more general feature representations, resulting in improved localization accuracy in the target environment. Finally, while transfer learning is highly beneficial in data-scarce target environments, its effectiveness diminishes when ample target data is available, as the target RMSE converges to a similar accuracy regardless of source dataset diversity.
\section{Limitations}   \label{sec:limit}
    While the dataset and underlying measurement campaign provide a unique combination of data type, amount of data, and precision, there are still limitations worth keeping in mind. 
    
    First and foremost, peculiarities of \ac{AP} characteristics are not eliminated from the CSI. The traditional way to solve this problem (and also to potentially include multi-antenna \acp{TX}) is the deployment of dedicated measurement \acp{TX}. This was not feasible for us since the cost for even just a few dozen such devices (so that characteristics between them and a given \ac{UE} location can be measured simultaneously) is prohibitive, as is the effort of deploying them in thousands of locations. Most importantly, such measurements would require an exclusive frequency license in the band of interest and/or the shutdown of the campus network for the duration of the measurement campaign (several months), neither of which is feasible. Thus, having uncalibrated \acp{TX} was the price to pay for large-scale data sets.
    
    Related to the use of \textcolor{black}{\ac{USC}} \acp{AP}, the exact location of the in-office \acp{AP} is prohibited from being disclosed by university rules. While such knowledge is not necessary for ML-based localization algorithms, it means that, e.g., \ac{ToA} based algorithms cannot be applied to our dataset.
    
    Secondly, even though our data set is by far the largest of its size, one can always wish for even more data. While we have $3,000$ \acp{AP}, one could always wish for $30,000$ or even $300,000$. Yet, given the time and effort of making precise measurements (our campaigns extended over more than 3 months, with 4-7 people involved in each measurement), this is not feasible for one research group. Still, it has to be kept in mind that the size of the dataset is limited, and while the buildings have different structures, they are still of the style (dimensions and material) typical for one specific city. 
    
    A third limitation of our measurements is that all locations are on straight lines, not on, e.g., a rectangular grid. Again, this follows naturally from the scaling of the measurement effort when multiple trajectories are to be measured, yet it might constitute a limitation for certain types of training and testing, and this should be kept in mind. 

    Fourth, While longer-term changes in the environment, e.g., new construction, can affect \ac{CSI}, measurements of such effects are beyond the scope of the current paper.
    
    Last but not least, the fact that we do perform measurements with a stable precision \ac{RX} while preventing the significant impact of human presence also implies that measurements with, e.g., smartphones held by a human user will show different characteristics, including variations between phone models and users holding the devices. Campaigns to measure human impact are worthwhile, but by definition, they are different from the scope of WiLoc. 
\vspace*{-0.2cm}
\section{Conclusion }    \label{sec:conclusion}
    This paper presented the results of a massive measurement campaign for location-labeled \ac{CSI} information designed for ML-based localization applications. This database, which is by far the largest of its kind, is made publicly available as part of this paper. We presented the measurement apparatus and procedure, verifications of the results, as well as measurement environments, sample results, and dataset format. Experiments with various ML localization algorithms demonstrated that the large size of the database in terms of \acp{AP}, \ac{UE} locations, and different buildings/environments allows to design, train, and/or test more accurate and/or robust ML algorithms. 

    Lastly, we note that our dataset can be used for ML applications beyond localization, such as channel prediction, modeling of handovers, etc. Investigations into specific schemes and their dependence on dataset size for those applications will be subject to our future work. \textcolor{black}{The authors have provided a public access link to their dataset: \\
    https://wides.usc.edu/programs-and-data.html\#wiloc. }
\vspace*{-0.2cm}
\section{Acknowledgment}
This work was supported in part by NSF projects CCF-2008443 and RINGS-2148315. The authors would like to thank the WiDeS group members for useful discussions and help with the measurements.

\bibliographystyle{IEEEtran}
\bibliography{References/reference}

\end{document}